# The quantum and classical Fano parameter *q*

**Masatomi Iizawa, Satoshi Kosugi, Fumihiro Koike and Yoshiro Azuma**

Department of Materials and Life Sciences, Sophia University, Tokyo 102-8554, Japan

Email: y-azuma@sophia.ac.jp

**Abstract**. The Fano resonance has been a familiar and important feature in atomic and molecular physics for more than half a century. Typically, the combination of a discrete state with one or more continua results in an asymmetric peak in the ionization spectrum. The peak-shape, called the Fano profile, can be expressed by a simple formula derived by Fano in 1935. However, the interpretation of its characteristic parameter $q$, which represents the asymmetry of the peak in the formula, is not necessarily intuitively transparent. The Fano resonance is not necessarily a quantum effect, but it is a manifestation of a certain physical mechanism in various systems, both quantum and classical. Through the derivation of $q$ from the known classical pictures with the classical coupled oscillator, we interpret the $q$ value with a geometrical view. And further, we introduce a complex valued $q$ parameter for the description of the resonance with a damped oscillator.

## 1. Introduction

Electronic excitation resonances found in atomic physics have been typically associated with symmetric peak-shapes called the Lorentzian, Breit-Wigner, or Cauchy distribution. However, in 1964, a very pronounced and asymmetric resonance was discovered in the helium photoabsorption spectrum due to photoexcitation to the doubly excited $2s2p$ $^1$P$^\circ$ state. It was the first experiment performed utilizing synchrotron radiation [1]. Nevertheless, the first theoretical formulation of this "Fano resonance" was developed sometime earlier by Fano (1935) [2] and was then extended and refined by Fano (1961) [3] comparing to electron impact excitation experiment data. (See also his surveys on atomic spectra [4], [5].) In the latter paper, it was argued that from the superposition state consisting of a discrete state $\varphi$ and a continuum $\psi_{E'}$

$$\Psi_E = a\varphi + \int dE' b_{E'}\psi_{E'} \quad (a \neq 0) \tag{1}$$

where $a$ and $b_{E'}$ are coefficients described by the functions of $E$ (eq. 2 in Fano (1961) [3]), the Fano profile formula for photoabsorption cross section $\sigma(\epsilon)$ is given by

$$\sigma(\epsilon) \propto \frac{(q+\epsilon)^2}{1+\epsilon^2} = 1 + \frac{q^2 - 1 + 2q\epsilon}{1+\epsilon^2} \tag{2}$$

where $\epsilon$ is the reduced photon energy and $q$ is the so-called Fano's asymmetry parameter, which represents the ratio between the photoexcitation amplitudes to the discrete state component and to the background continuum.

We show in figure 1, Fano's spectral profile curves eq. (2) for several values of the parameter $q$. When $q$ is close to zero, which is the case with vanishing discrete state excitation, the profile gives an almost symmetric window type resonance. When $|q|$ is large, which is the case with vanishing background continuum excitation, the profile gives an almost symmetric positive peak. For moderate values of $q$, the curves exhibit clear asymmetric peak profiles with the minimum value going down to zero at $\epsilon = -q$. (Note: If the state includes two or more continua, this spectrum does not necessarily go down to zero (section 4 in Fano (1961) [3]).)

This characteristic shape enables us to easily identify Fano resonances as conspicuous features in the experimental spectrum. Thus, it is interesting both in view of theory and practical utility. However, some of the important concepts in Fano's paper [3] are not necessarily obvious. These include the basic meaning of the $q$ parameter (also called the profile index or Fano parameter) or an explanation of why the peak becomes asymmetric when a bound state is superposed on a continuum. To illustrate the physical essence of the Fano resonance, we first discuss it in a classical system instead of a quantum system. Fano resonances are often regarded as a phenomenon specific to quantum mechanics [6], but they also occur in classical mechanics. See our Appendix for introduction to some typical examples.

A recent analytical study by Riffe (2011) [7] presented Fano resonances occurring in certain classical linear systems. However, Riffe's paper does not emphasize the fact that the Fano resonance is a ubiquitous phenomenon common to various physical systems rather than a concept valid only for a particular system. Some authors introduced the so-called "classical Fano resonance" as distinct from the (quantum) Fano resonance [9] [10] [11]. Also, a similar line shape in the classical coupled oscillator was noted [8]. They based the analogy in terms of similar peak-shapes. In the present paper, we discuss the nature of these Fano resonances in terms of the physical mechanism instead of just looking for analogy or metaphor in terms of the peak-shape. To demonstrate this, we analytically derive the Fano profile from the known classical model in the case of a coupled oscillator [9] [10] [11]. Through the derivation, we can define the Fano parameter $q$ with the coupled oscillator model.

It is known that there are also many other types of asymmetric spectral peaks that can be observed. An example would be a system that undergoes post-collision interactions after various atomic resonant ionizations [12]. (This interaction is also called final-state interaction in the context of nuclear and particle physics [13].) In those systems, the asymmetry is not due to interference between different ionization channels, but, by mutual interactions between two outgoing charged particles. It is important that Fano's profile formula can be derived from the process to verify that the spectral deformation is due to the interference within the system under consideration, thus constituting a Fano resonance.

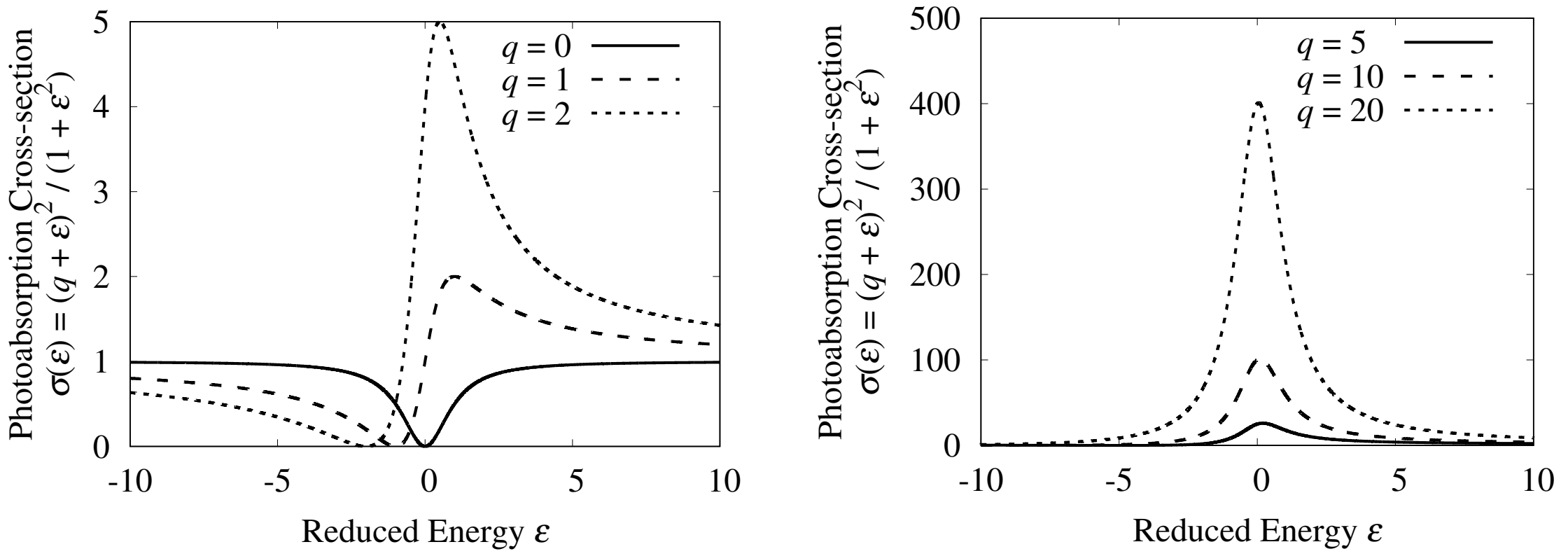

Figure 1. Graphical plots of Fano's formula for photoabsorption cross section $\sigma(\epsilon) = (q+\epsilon)^2/(1+\epsilon^2)$ for several values of the Fano parameter $q$, where $\epsilon$ is the reduced energy. When $q$ is large, the profile is almost symmetric. Otherwise, when $q$ is small, the profile is of an almost symmetric window type resonance.

## 2. Two weakly coupled harmonic oscillators with one of them damped

To discuss the interplay between a continuum oscillation and a discrete resonant oscillation, we consider the system of oscillators consisting of a damped resonance oscillator and a harmonic oscillator whose resonance frequency is overlapped by the broadened tail of the damped oscillation [9] [10] [11]. We illustrate the system in figure 2. We consider an oscillator $S_1$ with mass $m_1$, spring constant $k_1$, and damping coefficient $\gamma_1$, and another oscillator $S_2$ with mass $m_2$, spring constant $k_2$, and damping coefficient $\gamma_2$. The oscillators $S_1$ and $S_2$ are connected to fixed walls as illustrated. The masses $m_1$ and $m_2$ are connected by a weak spring with constant $K$ giving a weak coupling between the two oscillators $S_1$ and $S_2$. The system undergoes forced oscillations through the given external oscillatory force $F(t) = F_0 \cos(\omega t)$, where $F_0, \omega,$ and $t$ are amplitude, angular frequency, and time. When the damping coefficient $\gamma_1$ is large, the oscillator $S_1$ shows a broad resonance oscillation profile with the input of $F(t)$. We can observe the interaction of this broad resonance and the resonance of $S_2$.

Let us take $x_1(t)$ and $x_2(t)$ as the displacements of $m_1$ and $m_2$ from their equilibrium points.

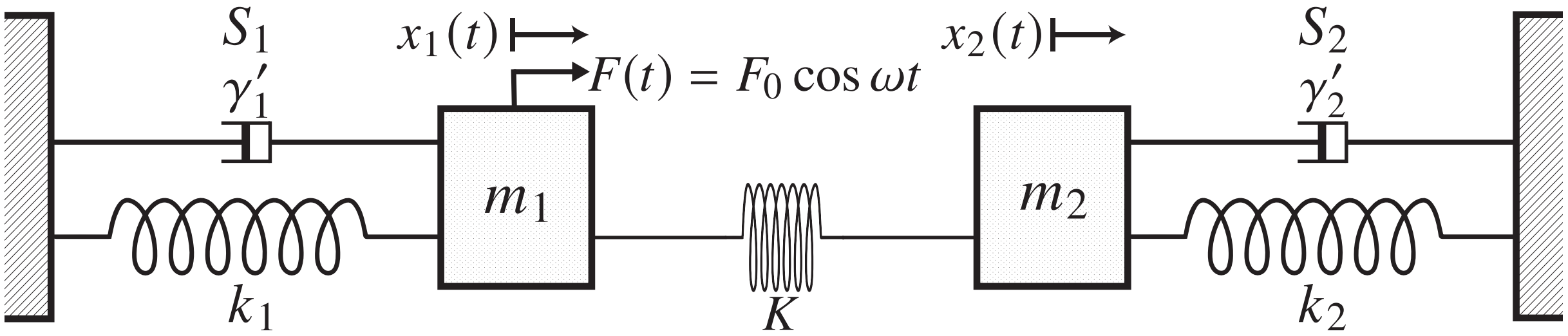

Figure 2. Damped coupled oscillator. $F(t)$ is the periodic external force for the oscillator $S_1$, $\gamma_1{}'$, $\gamma_2{}'$ are the dumping factors, $k_1, k_2$ are the spring constants, $K$ is the spring constants of the weak coupling, and $x_1, x_2$ are the displacements.

The equation of motion of this system is given by

$$\ddot{x}_1 + \gamma_1 \dot{x}_1 + \omega_1^2 x_1 + v_{12} x_2 = F_0 \cos(\omega t), \tag{3}$$

$$\ddot{x}_2 + \gamma_2 \dot{x}_2 + \omega_2^2 x_2 + v_{12} x_1 = 0\,, \tag{4}$$

where $\omega_1 = \sqrt{(k_1 + K)/\, m_1}$, $\omega_2 = \sqrt{(k_2 + K)/m_2}$, $\gamma_1 = \gamma_1'/m_1$, $\gamma_2 = \gamma_2'/m_2$ and $v_{12} = -K/m_1$. For simplicity of discussion, we call $\gamma_1$and $\gamma_2$ the damping factor of $S_1$ and $S_2$, and $v_{12}$ the coupling between $S_1$ and $S_2$, from now on. We can solve the equations (3) and (4) and obtain the steady state solution for $x_1(t)$. Let us take $c_1(\omega)$ as the amplitude of steady state oscillation $\cos(\omega t)$ and we assume the following form for $x_1(t)$, that is

$$x_1(t) = c_1(\omega) \cos(\omega t). \tag{5}$$

Substituting eq. (5) into eqs. (3) and (4), we obtain $c_1(\omega)$ as

$$c_1(\omega) = \frac{\omega_2^2 - \omega^2 + i\gamma_2\omega}{(\omega_1^2 - \omega^2 + i\gamma_1\omega)(\omega_2^2 - \omega^2 + i\gamma_2\omega) - v_{12}^2} F_0 \tag{6}$$

In the limit of vanishing coupling $v_{12} \to 0$, the amplitude $c_1(\omega)$ just becomes one independent simple damped oscillator, that is

$$c_1(\omega) = \frac{1}{(\omega_1^2 - \omega^2 + i\gamma_1\omega)} F_0 \tag{7}$$

The damped oscillation has, therefore, a phase shift $-\cot^{-1}((\omega^2 - \omega_1^2)/\gamma_1\omega)$ with respect to the external oscillatory force $F_0 \cos(\omega t)$. In the case of finite coupling $v_{12} = -K$, the amplitude $c_1(\omega)$ will give a characteristic resonance behavior reflecting the presence of the second oscillator $S_2$ near the frequency $\omega = \omega_2$. The phase shift at $\omega = \omega_2$, $-\cot^{-1}((\omega_2^2 - \omega_1^2)/\gamma_1\omega_2)$, of the simple damped oscillation, will behave as the background phase shift for the resonance behavior of $c_1(\omega)$ in the vicinity of $\omega = \omega_2$. Also, in the vicinity of $\omega = \omega_2$, the phase of the second oscillator $S_2$, $-\cot^{-1}((\omega^2 - \omega_2^2)/\gamma_2\omega)$ increases by $\pi$ when $\omega$ increases passed the value $\omega_2$ from below. When $\gamma_2$ is infinitesimally small, the change will take place within the immediate vicinity of $\omega = \omega_2$, giving the resonance oscillation. In the case of present coupled oscillators, this resonance oscillation takes place under the influence of a background phase shift $-\cot^{-1}((\omega_2^2 - \omega_1^2)/\gamma_1\omega_2)$, providing us with asymmetric resonance profiles; the background phase shift works as the starting phase of the resonance phase shift. In section 5, we discuss and graphically illustrate the role of those phase shift in relation to the Fano's $q$-parameter.

## 3. Interpretation of the Fano parameter $q$

In the last section, we reviewed and provided interpretation for previous discussions [9] [10] [11]. To observe the behavior of this system, we plot $|c_1(\omega)/F_0|^2$ against $\omega^2$ for a model case with several selected values of the damping coefficient $\gamma_1$ in figure 3. In a classical harmonic oscillator with $C$, energy $E$ can be given by

$$E = m\omega^2 C^2 = kC^2. \tag{8}$$

We find that the square of the frequency $\omega^2$ is linearly related to the energy $E$. To study the quantum analogue in the classical mechanical system, it is desirable to plot $|c_1(\omega)/F_0|^2$ against $\omega^2$ instead of $\omega$ itself, although it will be shown later in section 4 that the Fano profile formula in eq. (2) can naturally be derived as a function of $\omega$. We can recall that the distance of $\omega^2$ from a resonance point $\omega_r^2$, $\omega^2 - \omega_r^2 = (\omega + \omega_r)(\omega - \omega_r) \cong 2\omega_r(\omega - \omega_r)$, linearly depends on ω in the vicinity of $\omega^2 - \omega_r^2 \cong 0$, in other words, $\omega^2 - \omega_r^2$ is proportional to $\omega - \omega_r$ near the resonance point. In the quantum mechanical system, the energy $E$ depends linearly on the angular frequency $\omega$ according to Einstein's relation $E = \hbar\omega$, whereas $E$ is proportional to $\omega^2$ in the classical mechanical system as seen in eq. (8). We take $\omega^2$ as the variable to describe the profile formula for the convenience to compare the present system to the quantum mechanical systems.

In figures 3 to 5, we set the values of other parameters as $\omega_1 = 1.0$, $\omega_2 = 1.2$, $\gamma_2 = 0$, and $v_{12} = 0.1$. In figure 3, we present two resonance spectral profiles $|c_1(\omega)/F_0|^2$ (a) and (b) for the value $\gamma_1 = 0.025$ in as the model of weak damping. Figure 3(a) shows the overall feature of the spectral profile, while figure 3(b) gives an enlarged view near $\omega^2 = \omega_2^2$. In figures 4 and 5, we present the figures (a) and (b) similar to figure 3 for $\gamma_1 = 1.41$ and $\gamma_1 = 10.0$ as the models of moderate and strong damping, respectively. Looking at figure 3(a), we find two peaks. The peak to the left is the resonance profile due to the oscillator $S_1$ with intrinsic frequency $\omega_1$, and the peak to the right is due to the influence of the oscillator $S_2$ with intrinsic frequency $\omega_2$ realized by the coupling $v_{12}$ between $S_1$ and $S_2$. Let us call the latter profile $P_2$. In figure 3(b), we show the enlarged view of $P_2$. We can see that $P_2$ exhibits a characteristic asymmetric peak shape that is realized by the superposition of the tail of the damped oscillator $S_1$ and the simple oscillation at $\omega_2$. As found in eq. (7), this tail has a phase shift $-\cot^{-1}((\omega_2^2 - \omega_1^2)/\gamma_1\omega_2)$ in damped oscillation at the resonance point of $P_2$; the superposition of the delayed damped background oscillation and the intrinsic oscillation at $\omega_2$ provides us with an asymmetric profile. In the next section, we will show that this asymmetric profile is one of the so-called Fano profiles. Respectively, in figures 4(b) and 5(b), we can observe asymmetric profiles $P_2$ for larger attenuation coefficients $\gamma_1 = 1.41$ and $10.0$. For larger $\gamma_1$ the phase shift $-\cot^{-1}((\omega_2^2 - \omega_1^2)/\gamma_1\omega_2)$ becomes almost $\pi/2$, and the window type resonance profile emerges.

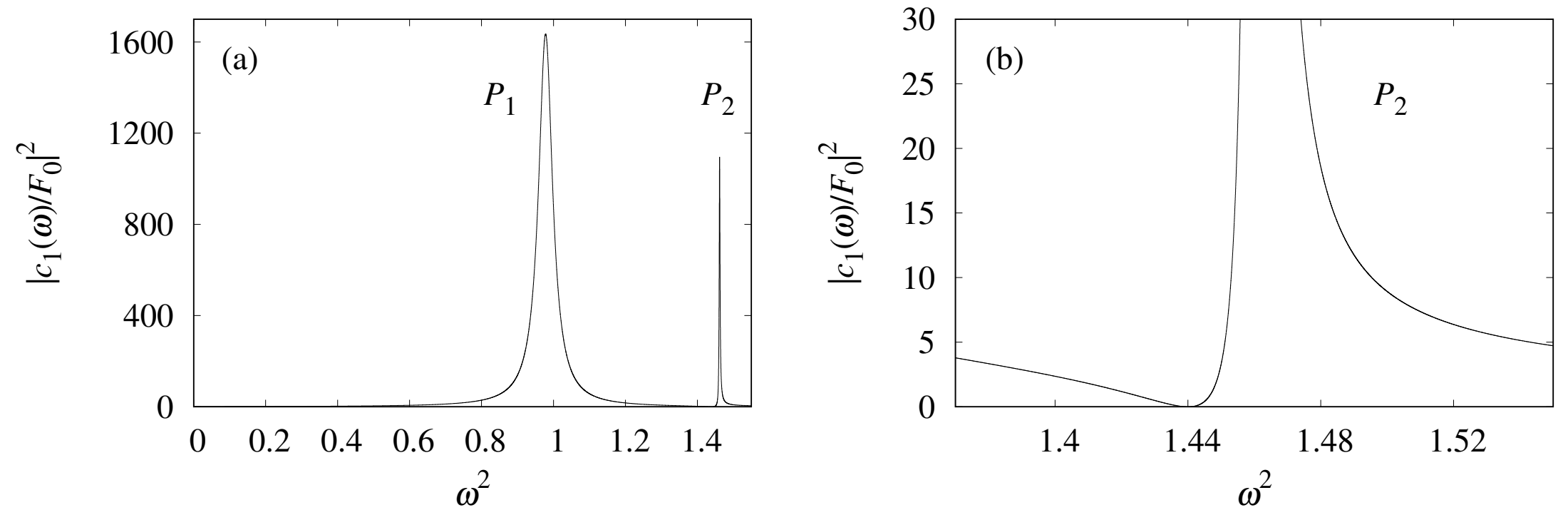

Figure 3. (a) The square of the amplitude of $x_1$ i.e., $|c_1(\omega)/F_0|$ vs. the square of the frequency of the external oscillatory force $\omega^2$ in case of the coupled oscillator indicated in Figure 2 with $\gamma_1 = 0.025, \gamma_2 = 0, v_{12} = 0.1, \omega_1 = 1.0, \omega_2 = 1.2$. (b) The enlarged plot around the right asymmetric peak of (a), called $P_2$.

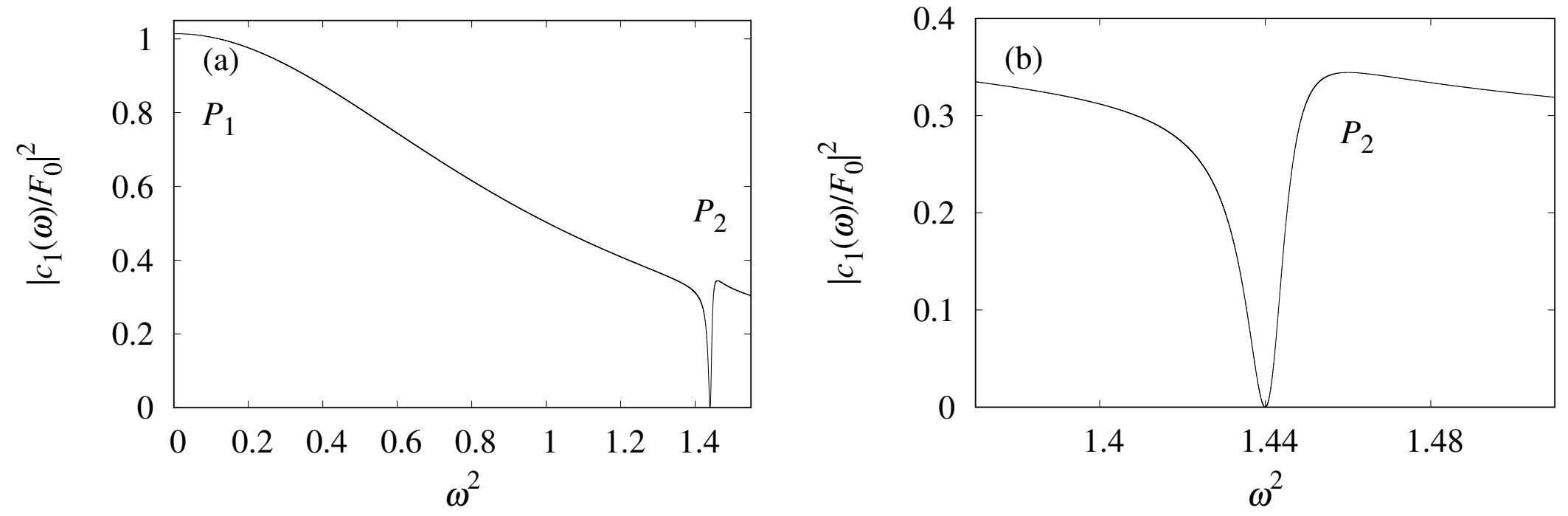

Figure 4. (a) The square of the amplitude of $x_1$ i.e., $|c_1(\omega)/F_0|$ vs. the square of the frequency of the external oscillatory force $\omega^2$ in case of the coupled oscillator indicated in Figure 2 with $\gamma_1 = 1.41$, $\gamma_2 = 0, v_{12} = 0.1, \omega_1 = 1.0, \omega_2 = 1.2$. (b) The enlarged plot around $P_2$.

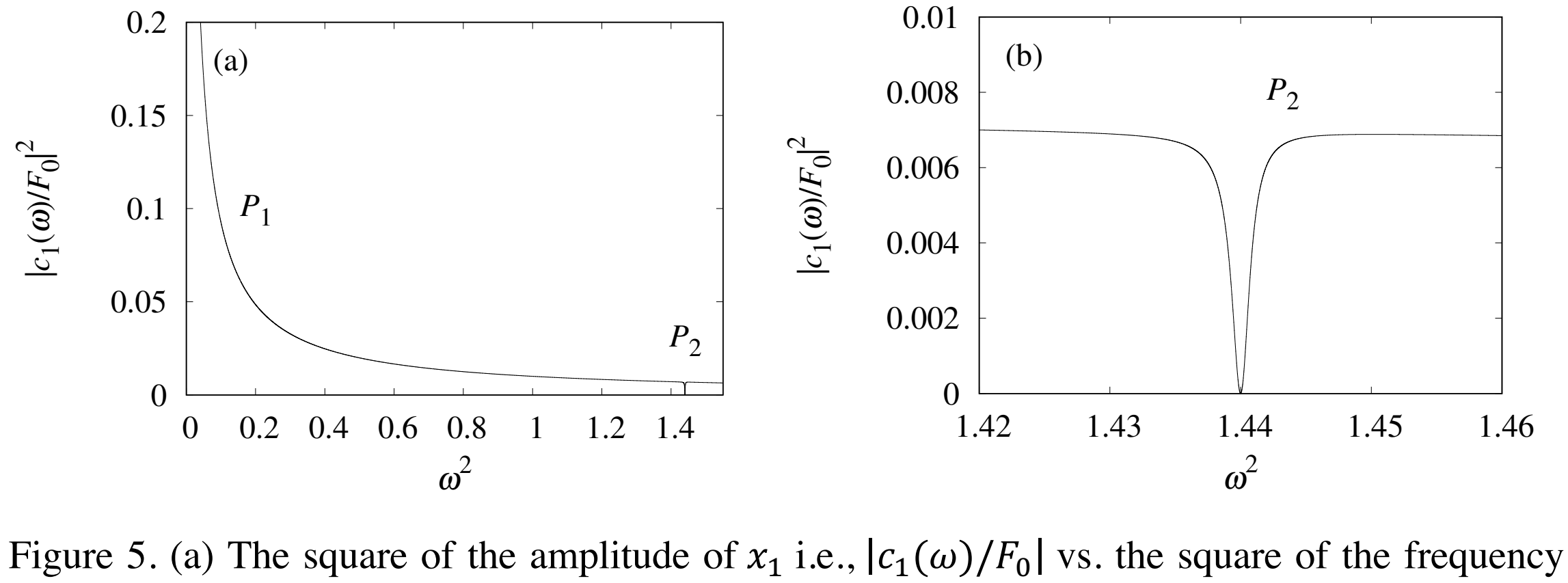

Figure 5. (a) The square of the amplitude of $x_1$ i.e., $|c_1(\omega)/F_0|$ vs. the square of the frequency of the external oscillatory force $\omega^2$ in case of the coupled oscillator indicated in Figure 2 with $\gamma_1 = 10$,

$\gamma_2 = 0, v_{12} = 0.1, \omega_1 = 1.0, \omega_2 = 1.2$. (b) The enlarged plot around $P_2$. It looks like a symmetric dip. This dip is called the *window resonance* [14].

## 4. Derivation of the Fano formula for the classical coupled oscillator

In the present section, we investigate the magnitude of the oscillation of the system given in figure 3-5, $|c_1(\omega)|^2$ , in detail for the vicinity of $\omega^2 = \omega_2^2$, and try to theoretically confirm that the profile $P_2$ introduced in the previous section is the Fano profile. Now, we calculate

$$|c_1(\omega)|^2 = c_1(\omega)c_1^*(\omega), \tag{9}$$

where the asterisk * means the complex conjugate, in the vicinity of the profile $P_2$ under the condition $\gamma_2 = 0$ as in the previous section.

When diagonalizing the equations (3) and (4), it can be seen that the resonance frequencies are somewhat shifted from the intrinsic frequencies $\omega_1, \omega_2$. We can observe in figure 4(a), for example, the slight deviations of the profiles from their natural resonance positions $\omega^2 = \omega_1^2$ or $\omega^2 = \omega_2^2$. We find that

$$\omega_a^2 = \omega_1^2 - \frac{v_{12}^2}{\omega_2^2 - \omega_1^2}, \tag{10}$$

$$\omega_b^2 = \omega_2^2 + \frac{v_{12}^2}{\omega_2^2 - \omega_1^2}, \tag{11}$$

are the new resonance position, which is shifted by the same amount $v_{12}^2/(\omega_2^2 - \omega_1^2)$ in the opposite directions due to the coupling $v_{12}^2$. We define the "energy" of the system $\tilde{\epsilon}$ as

$$\tilde{\epsilon} \equiv \omega^2 - \omega_b^2 = \omega^2 - \omega_2^2 - \frac{v_{12}^2}{\omega_2^2 - \omega_1^2}. \tag{12}$$

In equation (12), since $\tilde{\epsilon} = \omega^2 - \omega_b^2 = (\omega + \omega_\mathrm{b})(\omega - \omega_\mathrm{b})$, we can regard $(\omega + \omega_\mathrm{b})$ as a constant if we consider it in the narrow neighborhood of $\tilde{\epsilon} = 0$. Therefore, since $\tilde{\epsilon} \propto (\omega - \omega_\mathrm{b})$, the same argument would hold for the energy even though the first power of the frequency was considered.

We take $c_1(\omega) = (A/B)F_0$ in eq. (6). Then, under the condition that $\gamma_2 = 0$, we have, for the numerator $A$,

$$\begin{aligned} A &= \omega_2^2 - \omega^2 = -\tilde{\epsilon} - \frac{v_{12}^2}{\omega_2^2 - \omega_1^2} \\ &= -\gamma_1\omega_2 \frac{v_{12}^2}{(\omega_2^2 - \omega_1^2)^2}\left[\frac{1}{\gamma_1\omega_2}\frac{(\omega_2^2 - \omega_1^2)^2}{v_{12}^2}\tilde{\epsilon} + \frac{1}{\gamma_1\omega_2}(\omega_2^2 - \omega_1^2)\right], \end{aligned} \tag{13}$$

and we have, for the denominator $B$,

$$\begin{aligned} B &= (\omega_1^2 - \omega^2 + i\gamma_1\omega)(\omega_2^2 - \omega^2) - v_{12}^2 \\ &= (\omega_2^2 - \omega_1^2)\tilde{\epsilon} - i\gamma_1\omega_2 \frac{v_{12}^2}{\omega_2^2 - \omega_1^2}, \end{aligned} \tag{14}$$

using the approximation that

$$\frac{v_{12}^2}{\omega_2^2-\omega_1^2} \gg \tilde{\epsilon} \approx 0 \tag{15}$$

in the neighborhood of $\tilde{\epsilon} = 0$. This is the only approximation employed in this derivation. Summarizing the above, the amplitude $c_1(\omega) = (A/B)F_0$ can be written as

$$c_1(\omega) \approx \frac{-\gamma_1\omega_2 \frac{v_{12}^2}{(\omega_2^2-\omega_1^2)^2}\left[\frac{1}{\gamma_1\omega_2}\frac{(\omega_2^2-\omega_1^2)^2}{v_{12}^2}\tilde{\epsilon} + \frac{1}{\gamma_1\omega_2}(\omega_2^2-\omega_1^2)\right]}{\gamma_1\omega_2\frac{v_{12}^2}{\omega_2^2-\omega_1^2}\left[\frac{1}{\gamma_1\omega_2}\frac{(\omega_2^2-\omega_1^2)^2}{v_{12}^2}\tilde{\epsilon} - i\right]} F_0 \tag{16}$$

Looking closely into equation (16), certain expressions can be observed. One of them is $\frac{1}{\gamma_1\omega_2}\frac{(\omega_2^2-\omega_1^2)^2}{v_{12}^2}\tilde{\epsilon}$, then we can define the reduced energy $\epsilon$ by

$$\epsilon \equiv \frac{1}{\gamma_1\omega_2}\frac{(\omega_2^2-\omega_1^2)^2}{v_{12}^2}\tilde{\epsilon}. \tag{17}$$

Further, we define a parameter $q$ by

$$q \equiv \frac{1}{\gamma_1\omega_2}(\omega_2^2-\omega_1^2). \tag{18}$$

Then, the amplitude equation (16) can be written as

$$c_1(\omega) \approx -\frac{\epsilon+q}{\epsilon-i}\frac{F_0}{\omega_2^2-\omega_1^2}. \tag{19}$$

Therefore, $|c_1(\omega)|^2$ can be written as

$$|c_1(\omega)|^2 \approx \frac{(\epsilon+q)^2}{\epsilon^2+1}\frac{F_0^2}{(\omega_2^2-\omega_1^2)^2} \propto \frac{(\epsilon+q)^2}{\epsilon^2+1}, \tag{20}$$

since $F_0^2/(\omega_2^2-\omega_1^2)^2$ is constant. This is nothing but the Fano's profile formula itself. That means that $q$ is the Fano's profile parameter, which produces the figure 2.

Comparing eq.(21) and eq.(18), we find that the phase shift of the damped oscillator at $\omega = \omega_2$, $-\cot^{-1}((\omega_2^2-\omega_1^2)/\gamma_1\omega_2)$, is equal to $-\cot^{-1} q$, giving the degree of the asymmetry of the peak profile near $\omega = \omega_2$. And further, we find that the Fano parameter $q$ is inversely proportional to the attenuation coefficient $\gamma_1$. For large $\gamma_1$, $q \to 0$, and then, the phase shift goes to $\to \pi/2$. The resonance oscillation near $\omega = \omega_2$ is modified by the background damped oscillation providing us with a window type resonance profile [14].

## 5. Fano resonance projected on a plane tangential to the cylindrical coordinate surface

To understand the Fano parameter $q$ defined in eq. (18), we try to investigate the resonance profile as a function of phase angles. We introduce an angle variable $\delta_\epsilon = -\cot^{-1}\epsilon$, and an angle parameter $\delta_q = -\cot^{-1} q$. Then, we can write the Fano's profile function as

$$\frac{1}{(1+q^2)}\frac{(q+\epsilon)^2}{1+\epsilon^2}=\sin^2\!\left(\delta_\epsilon+\ \delta_q\right). \qquad (21)$$

Here, we have introduced a factor $1/(1+q^2)$ to normalize the profile function $(q+\epsilon)^2/(1+\epsilon^2)$; in the limit of $q\ \rightarrow\ \pm\infty$ , in other words, in the limit of $\delta_q\ \rightarrow 0$, the left-hand side of the above equation gives a simple Lorentzian function $1/(1+\epsilon^2)$.

As discussed in section 2, the phase shift or phase delay of the damped oscillator $S_1$ at $\omega=\omega_2, -\cot^{-1}((\omega_2^2-\omega_1^2)/\gamma_1\omega_2)$ becomes the initial phase from which the resonance phase jump by $\pi$ at the resonance position of a simple oscillator $S_2$. Also, as introduced in eq. (18) of section 4, this phase shift is related to the Fano's asymmetry parameter $q$. In the present section we defined the phase $\delta_q$ as $\delta_q=\ -\cot^{-1}q=\ -\cot^{-1}((\omega_2^2-\omega_1^2)/\gamma_1\omega_2)$. Therefore, we can investigate the asymmetric resonance profiles in the coordinate of phase angles. The resonance can be represented by a simple rotation concerning this phase angle, and we can further see that $\delta_q$ is its initial phase.

Let us take $\delta=\delta_\epsilon+\ \delta_q$. We try to draw the profile function $\sin^2(\delta=\delta_\epsilon+\ \delta_q)$ given in eq.(21) on a cylindrical coordinate in figure 6. We further consider the projection of the profile function onto a plane which is tangential to the cylindrical coordinate surface. Viewing from the cylindrical coordinate origin, we find a symmetric Lorentzian profile ranging from $\delta=\ 0$ to $\pi$ on a plane which is tangential to the cylindrical coordinate surface at $\delta=\ \pi/2$. And, further on, we find an asymmetric profile curve on a plane tangential at $\delta\neq\ \pi/2$, which can be attributed to the Fano profile as discussed in detail below.

We, first, consider the case of $\delta_q\ =0$, which is the case of $q=\ -\cot\delta_q=\ -\cot 0=\ \pm\infty$. We plot $\sin^2(\delta=\delta_\epsilon+\ \delta_q)=\ \sin^2(\delta=\delta_\epsilon)$ on a cylindrical coordinate surface with radius 1 as illustrated in figure 7. And we project the curve from the origin of the cylindrical coordinate onto a plane which is tangential to the cylindrical coordinate surface at $\delta\ =\ \ \delta_\epsilon=\pi/2$, namely, at $\epsilon=-\cot\delta_\epsilon=\ -\cot(\pi/2)\ =0$. As shown in figure 7, the angle $\delta_\epsilon$ is projected onto $\varepsilon=\ -\cot\delta_\epsilon$ on the plane, and the projected curve represents the Lorentzian curve when we vary $\delta=\delta_\epsilon$ from 0 to $\pi$. Second, we consider the case of $\delta_q\ \neq 0$ and discuss the role of the phase $\delta_q$ in the resonance profile. We plot $\sin^2\!\left(\delta=\delta_\epsilon+\ \delta_q\right)$ on a cylindrical coordinate surface with radius 1 as shown in figure 8. We then project the curve from the origin of the cylindrical coordinate onto a plane which is tangential to the cylindrical coordinate surface at $-\ \delta_q=\ \ \delta_\epsilon=\pi/2$, namely, at $\varepsilon=\ -\cot\delta_\epsilon=\ -\cot(\pi/2)\ =0$. The projected curve represents the Fano profile curve when we vary $\delta_\epsilon$ from 0 to $\pi$, i. e. $\delta=\delta_\epsilon+\ \delta_q$ from $\delta_q$ to $\delta_q+\ \pi$. As an example, we have shown the profile curve for the case of $\delta_q\ =\ 3\pi/4$, which corresponds to the case of $q=1$. We can see an asymmetric profile on the tangential plane. The asymmetry is realized by the initial phase $\delta_q$ in the profile function $\sin^2\delta$. In other words, the Fano's parameter $q=\ -\cot\delta_q$ represents the initial phase $\delta_q$ of the profile function given in terms of the angle variable $\delta$. In the present paper, we are considering a coupled two harmonic oscillators with damping. The initial phase $\delta_q\ =\ -\cot^{-1}q=\ -\cot^{-1}((\omega_2^2-\ \omega_1^2)/\gamma_1\,\omega_2)$ of the resonance profile is the phase shift of the damped oscillation at $\omega=\ \omega_2$, as discussed in section 2. In the vicinity of the resonance point $\omega=\ \omega_2$, the angle variable $\delta=\delta_\epsilon+\ \delta_q$ of the resonance profile $\sin^2\delta$ changes by $\pi$ and $\delta_q$ gives its initial phase.

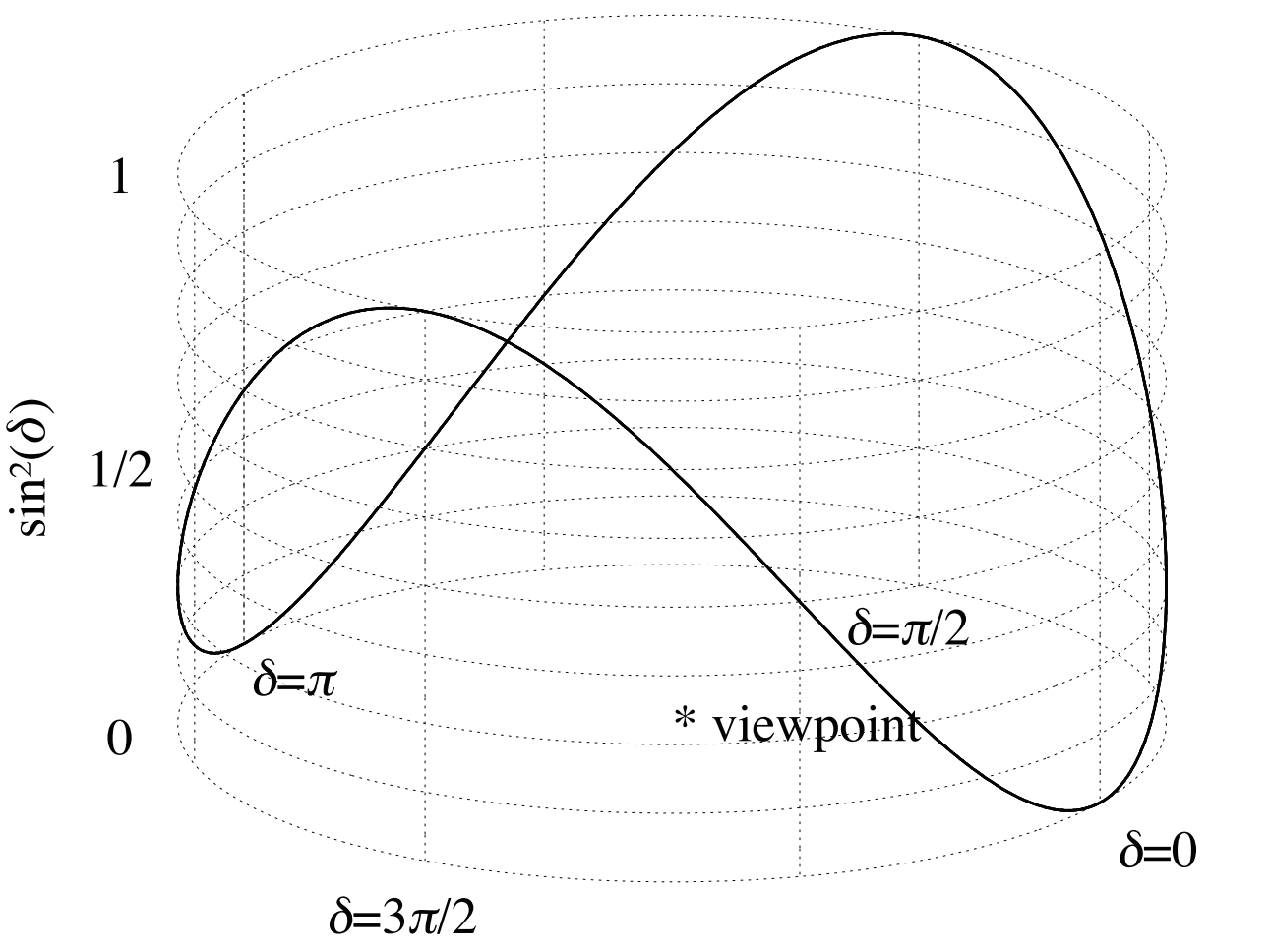

Figure 6. The profile function $\sin^2(\delta = \delta_\epsilon + \delta_q)$ on cylindrical coordinate with an angle variable $\delta_\epsilon = -\cot^{-1}\epsilon$, and an angle parameter $\delta_q = -\cot^{-1} q$. When $\delta_q \to 0$, namely $q \to \pm\infty$, then the profile function should be $\sin^2(\delta = \delta_\epsilon) = 1/(1+\epsilon^2)$, that is, or Lorentzian profile.

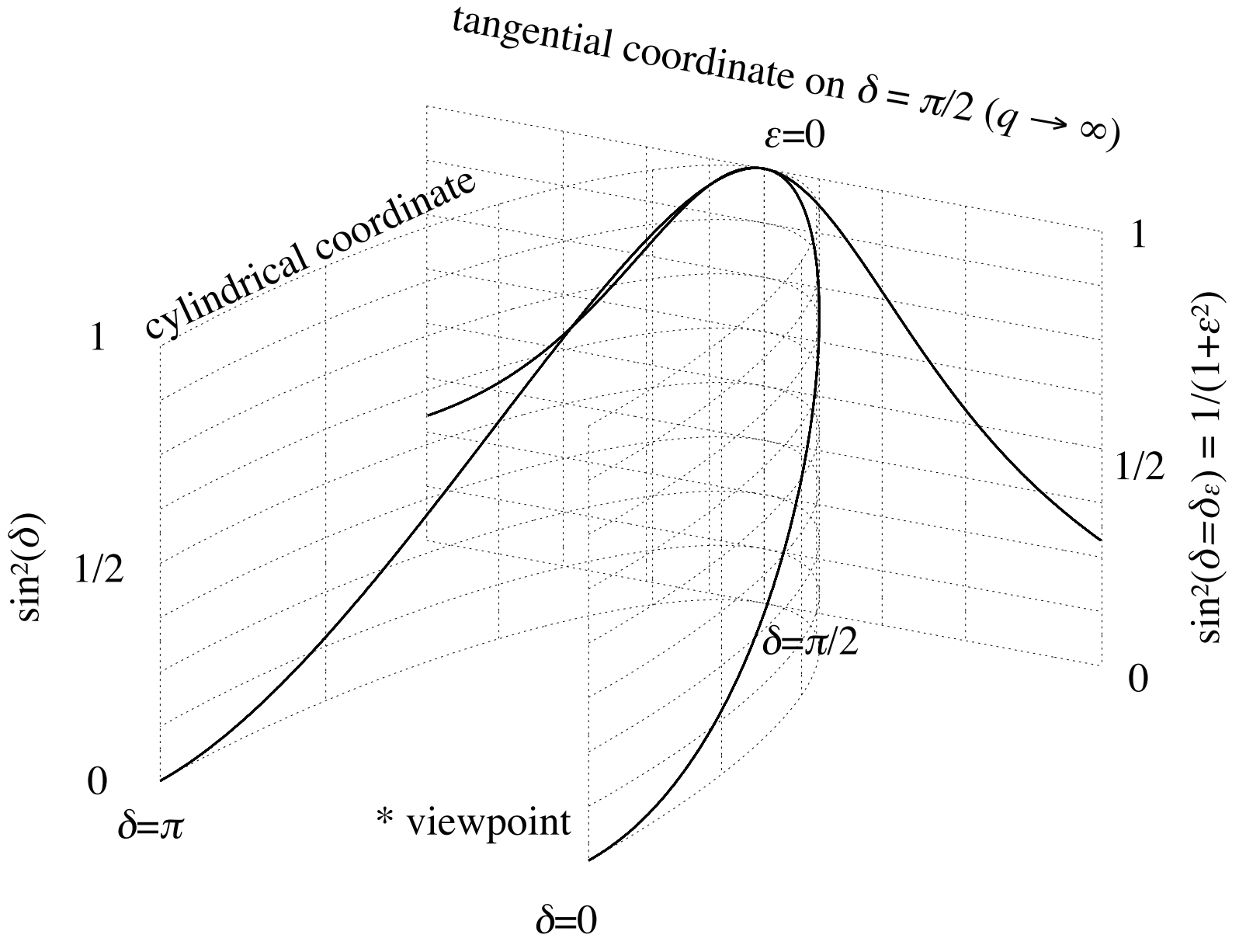

Figure 7. $\sin^2(\delta = \delta_\epsilon) = 1/(1+\epsilon^2)$ on a cylindrical coordinate surface with radius 1 in case of $\delta_q \to 0$, or $q \to \pm\infty$, that is, Lorentzian profile, the special case of the Fano profile in case of $q \to \pm\infty$. The plot on the cylindrical coordinate is tangential from the Lorentzian profile on the plane coordinate at $\delta = \delta_\epsilon = \pi/2$, namely, at $\epsilon = -\cot\delta_\epsilon = -\cot(\pi/2) = 0$.

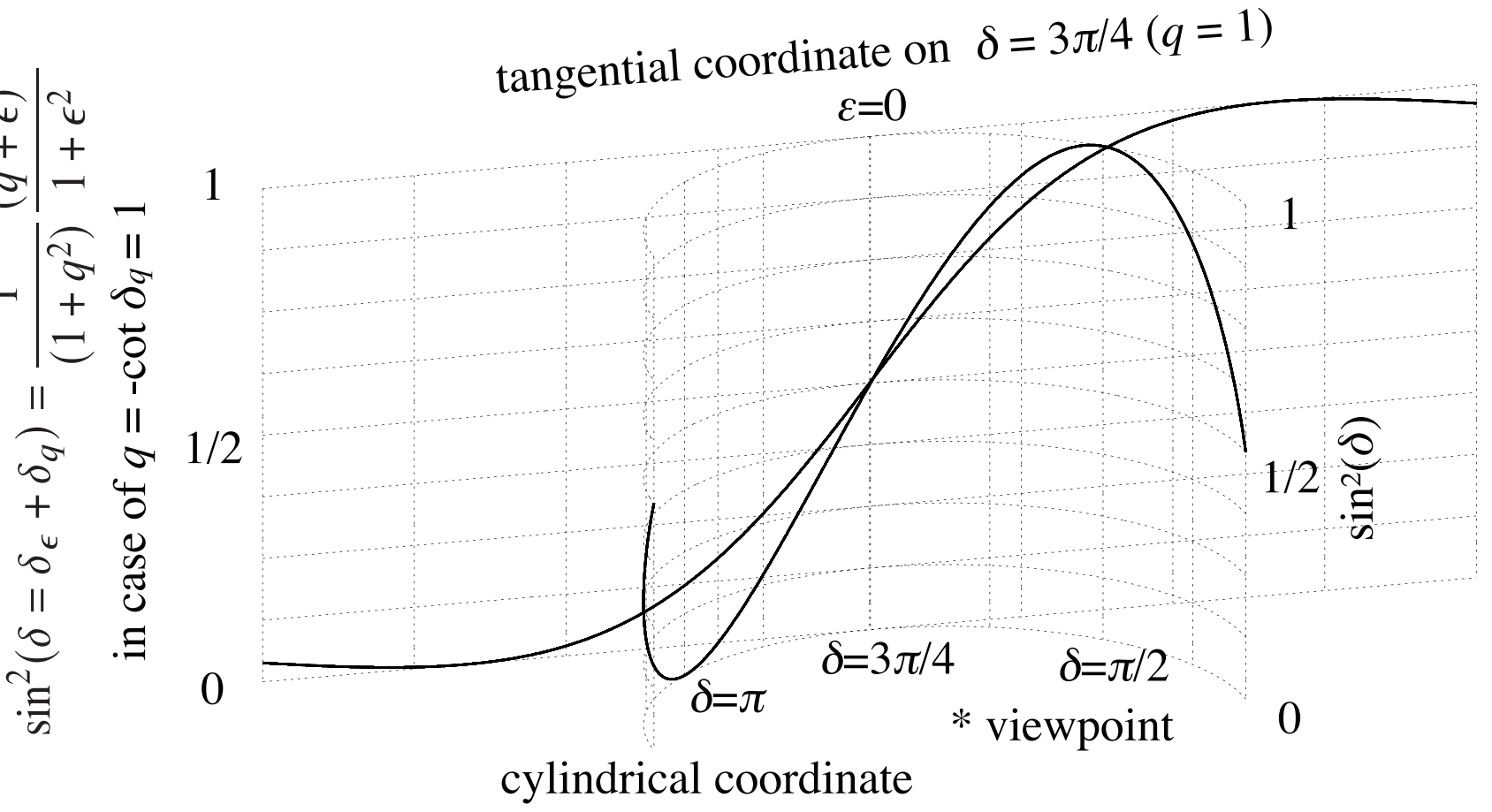

Figure 8. $\sin^2(\delta_\epsilon + \delta_q) = (1/(1+q^2))((q+\epsilon)^2/(1+\epsilon^2))$ on a cylindrical coordinate surface with radius 1 in case of $\delta_q = 3\pi/4$, or the Fano parameter $q = -\cot\delta_q = 1$. We can see an asymmetric profile on the tangential plane. The asymmetry is realized by the initial phase $\delta_q$ in the profile function $\sin^2\delta$.

## 6. Extension of the *q*-parameter for the case of a resonator with energy dissipation

In the present section, we try to consider the case that the damping factor $\gamma_2$ of oscillator $S_2$ is finite and positive, i.e. the case that we have a resonator $S_2$ with energy dissipation. As has been shown in the previous sections, a simple oscillator $S_2$ with $\gamma_2 = 0$ can be discussed in analogy with an isolated atom with a single discrete state. In atomic systems, a continuum that couples with a single atomic discrete state can undergo a complete interference between the two and have a chance to hit a zero point in the spectra at $\epsilon = -q$ in eq. (20). As we can see in Figs. 3-5, this is also the case for a damped oscillator $S_1$ coupled to an oscillator $S_2$ with a vanishing damping factor $\gamma_2$. For finite $\gamma_2$, we are to consider the system that has an interaction with the environment, or, in other words, a coupling with other channels. So, the complete cancelation of oscillations may not be possible, which leads to the chance for the spectral profile to hit zero being lost. In Figs. 9-11, we show examples of such spectral profiles.

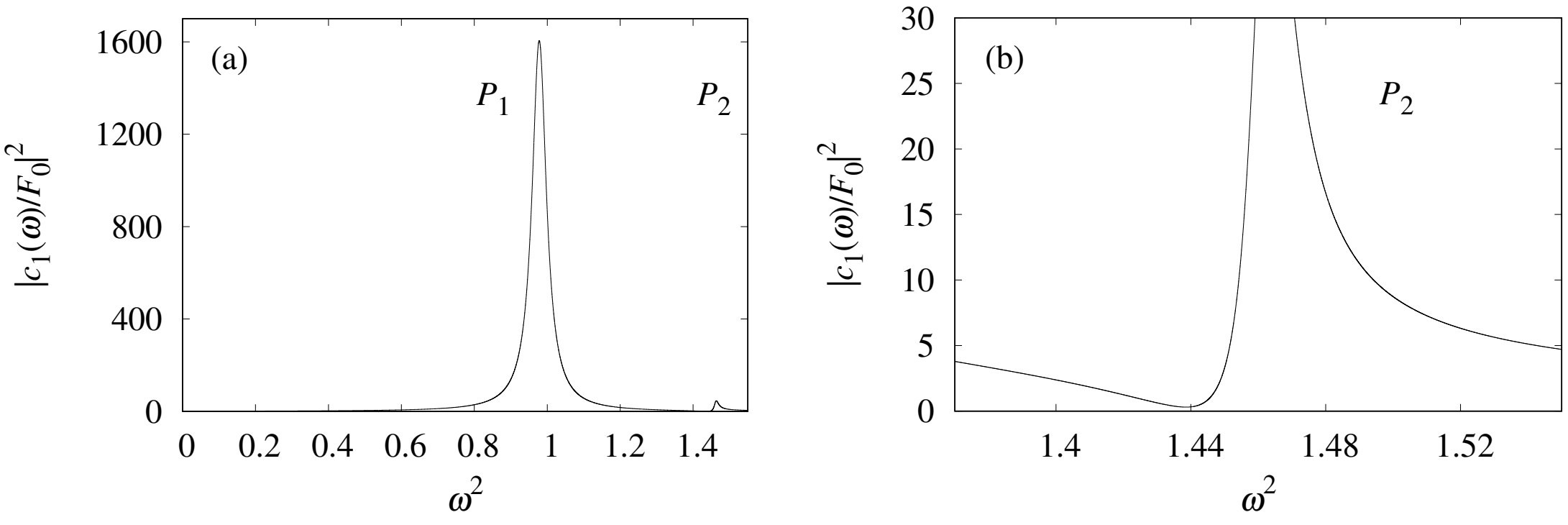

Figure 9. (a) The square of the amplitude of $x_1$ i.e., $|c_1(\omega)/F_0|$ vs. the square of the frequency of the external oscillatory force $\omega^2$ in case of the coupled oscillator indicated in Figure 2 with $\gamma_1 = 0.025, \gamma_2 = 0.005, v_{12} = 0.1, \omega_1 = 1.0, \omega_2 = 1.2$. The conditions are the same as in Figure 3 except

for $\gamma_2$. (b) The enlarged plot around the right asymmetric peak of (a), called $P_2$. Note that $|c_1(\omega)/F_0|$ doesn't go to zero at $\omega^2 = {\omega_2}^2 = 1.44$ .

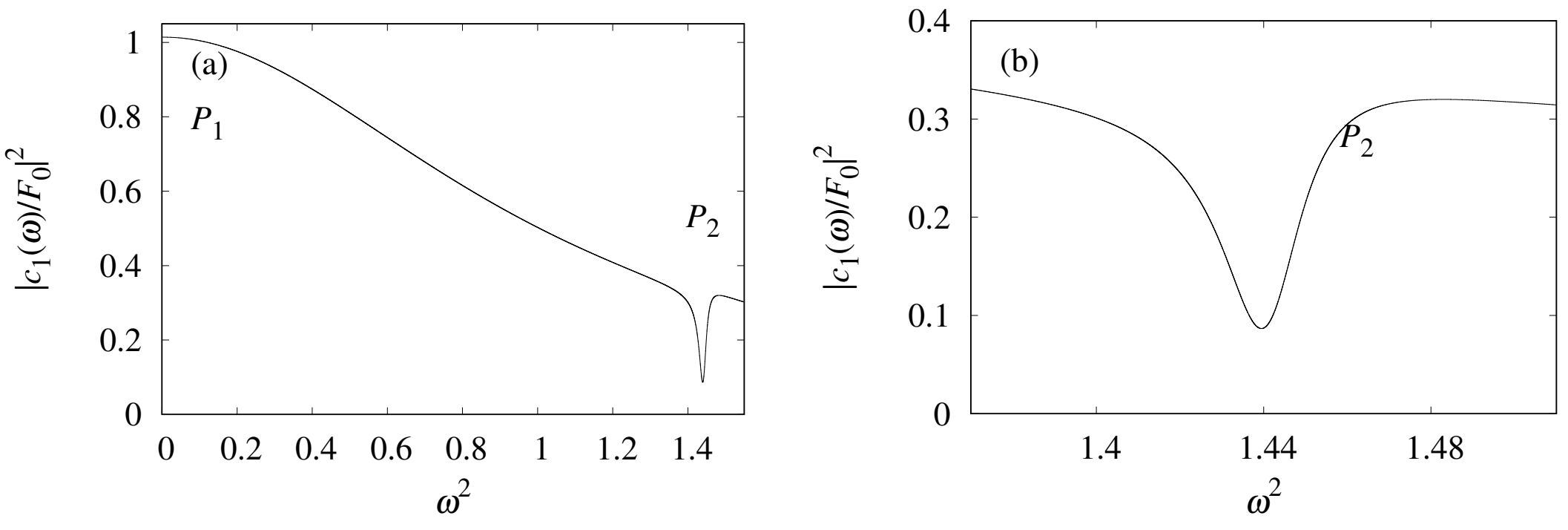

Figure 10. (a) The square of the amplitude of $x_1$ i.e., $|c_1(\omega)/F_0|$ vs. the square of the frequency of the external oscillatory force $\omega^2$ in case of the coupled oscillator indicated in Figure 2 with $\gamma_1 = 1.41$, $\gamma_2 = 0.005, v_{12} = 0.1, \omega_1 = 1.0, \omega_2 = 1.2$. The conditions are the same as in Figure 4 except for $\gamma_2$. (b) The enlarged plot around $P_2$. Note that $|c_1(\omega)/F_0|$ doesn't go to zero at $\omega^2 = {\omega_2}^2 = 1.44$ .

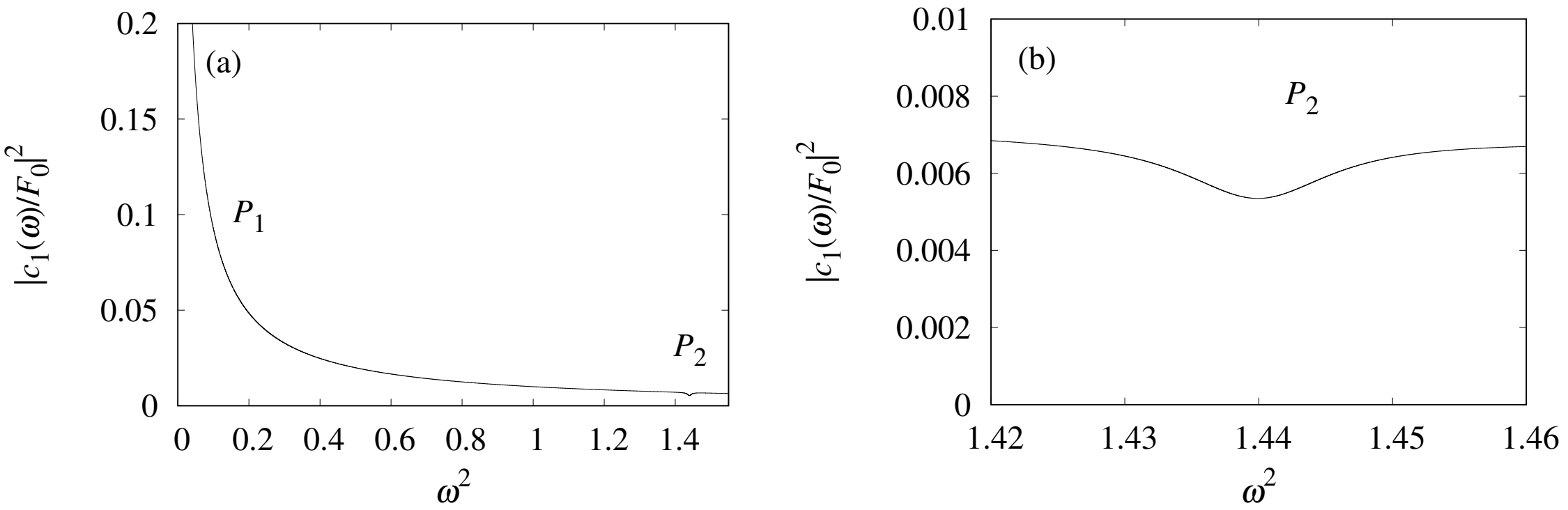

Figure 11. (a) The square of the amplitude of $x_1$ i.e., $|c_1(\omega)/F_0|$ vs. the square of the frequency of the external oscillatory force $\omega^2$ in case of the coupled oscillators indicated in Figure 2 with $\gamma_1 = 10$, $\gamma_2 = 0.005, v_{12} = 0.1, \omega_1 = 1.0, \omega_2 = 1.2$. The conditions are the same as in Figure 5 except for $\gamma_2$. (b) The enlarged plot around $P_2$. Note that $|c_1(\omega)/F_0|$ doesn't go to zero at $\omega^2 = {\omega_2}^2 = 1.44$ .

Now, we can show as follows that the effect of energy dissipation can be taken into account by introducing a complex valued Fano's $q$-parameter. In the case of $\gamma_2 \neq 0$, the numerator $A$ and denominator $B$ of the amplitude in eq. (6) $c_1(\omega) = (A/B)F_0$ are given as

$$A = \omega_2^2 - \omega^2 + i\gamma_2\omega \ , \tag{22}$$

and

$$B = (\omega_1^2 - \omega^2 + i\gamma_1\omega)(\omega_2^2 - \omega^2) - v_{12}^2 + (\omega_1^2 - \omega^2 + i\gamma_1\omega)i\gamma_2\omega \ . \tag{23}$$

We investigate the behavior of the amplitude $c_1(\omega)$ in the vicinity of $\omega = \omega_2$ . We can put $i\gamma_1\omega = i\gamma_1\omega_2$, and $i\gamma_2\omega = i\gamma_2\omega_2$ in eqs. (22) and (23). Further on, we consider the case of small damping factor $\gamma_2$ . We may assume that $(\omega_1^2 - \omega^2 + i\gamma_1\omega)i\gamma_2\omega = 0$ . Then we obtain the same expression eq. (19) for the amplitude $c_1(\omega)$ with the definition of $q$ as

$$q \equiv q_r + iq_i \equiv \frac{\omega_2^2 - \omega_1^2}{\gamma_1\omega_2}\left(1 - i\gamma_2\omega_2\frac{\omega_2^2 - \omega_1^2}{v_{12}^2}\right), \tag{24}$$

where, $q_r$ is the real part of $q$ which is the same as that of $q$ in eq. (18), and $q_i$ is the imaginary part of $q$, which is given by

$$q_i = -\frac{\gamma_2}{\gamma_1}\cdot\frac{(\omega_2^2 - \omega_1^2)^2}{v_{12}^2}. \tag{25}$$

In the limit of $\gamma_2 \to 0$, we have $q_i \to 0$, and the complex valued $q$ in eq. (24) results in a real valued $q$ defined in eq. (18).

Now, we have

$$|c_1(\omega)|^2 \approx \frac{(\epsilon+q)(\epsilon+q^*)}{\epsilon^2+1}\cdot\frac{F_0^2}{(\omega_2^2-\omega_1^2)^2} \propto \frac{(\epsilon+q)(\epsilon+q^*)}{\epsilon^2+1}, \tag{26}$$

where $q^*$ is the complex conjugate of $q$. Then, substituting $q = q_r + iq_i$ into eq. (26), we have

$$|c_1(\omega)|^2 \approx \frac{(\epsilon+q_r)^2+q_i^2}{\epsilon^2+1}\cdot\frac{F_0^2}{(\omega_2^2-\omega_1^2)^2}. \tag{27}$$

From this equation, we can see that the spectral profile does not go down to zero throughout the entire region of the reduced energy $\epsilon$. We can say that this is one of the characteristics of the energetically open system.

Note that the previous discussion on the complex Fano parameter in atomic physics was based on quantum scattering theory [15], but our discussion has made it possible to extend it regardless of quantum or classical theory.

## 7. Conclusion

Fano's resonance profile formula has been derived for the forced oscillation of coupled classical oscillators. Asymmetric resonance profiles following the Fano's profile formula have been observed in such systems. For the case $\gamma_2 = 0$, i.e., $S_2$ is an undamped simple oscillator, the Fano's profile parameter $q$ is given by $(\omega_2^2 - \omega_1^2)/(\gamma_1\omega_2)$ as in eq.(18), and the corresponding angle parameter $\delta_q = -\cot^{-1} q = -\cot^{-1}((\omega_2^2 - \omega_1^2)/(\gamma_1\omega_2))$ is the phase shift in the forced simple oscillation with damping coefficient $\gamma_1$. When the energy dissipation is weak, i.e., $\gamma_1 \to 0$, $\delta_q$ is almost zero and $q$ is quite large, providing us with almost a symmetric resonance profile. On the other hand, when $\gamma_1$ is large, $\delta_q$ is almost $\pi/2$ and $q$ is almost zero, providing us with a window type resonance profile. The phase shift represented by the angle parameter $\delta_q$ can be considered as an analogue to the background phase shift in quantum mechanical atomic resonance scattering processes.

A damping factor $\gamma_2$ with a finite value leads to a Fano's profile with a complex valued $q$-parameter. Further investigations for the Fano resonance oscillations of the open system will be pursued in a future publication.

## Acknowledgments

We were indebted to Prof. A. R. P. Rau (Louisiana State University) for constructive comments. He introduced us to his work on the general expression on continuum-discrete coupling [16] which was helpful for the improvement of our early manuscript. We also thank Dr. Yasuhito Narita (Austrian Academy of Sciences) for insightful comments and suggestions.

## Appendix. Examples of classical Fano resonance

The fact that a Fano resonance can occur in the system of spring and mass means that Fano resonance can be found in various places in everywhere and everyday life. The immediately imaginable one is the Fano resonance using the RLC resonance circuit [10][11]. For every article cited as references, they simply pointed out that as a result of plotting the voltage gain against the frequency of the input voltage for the LCR circuit equivalent to the coupled oscillator, the peak becomes asymmetric and they assert that it is a metaphor of Fano resonance. Given the series of results in our paper, these are not "metaphor," but it can be said that Fano resonance is occurring because the Fano profile can be derived analytically.

Also, in acoustics, there is a precedent for Fano resonance by resonance tubes and the similar systems [17] [18] [19] [20] [21]. There is another interesting example in optical engineering which attempts to switch light of a specific frequency at high speed by changing the Fano parameter [22]. This example in optical engineering is easy to understand by an analogy with the waveguide reduced to a nanoscale. When the electromagnetic wave leaks out of the waveguide from the slight gap at the joint of the waveguide (since the frequency characteristic is flat outside because it is outside the tube, it will be going through the continuum) and then overlaps with the wave that has passed through the waveguide (It can be said that it went through a discrete state system because it forms a sharp Lorentzian only at specific frequencies) at the end, a Fano resonance occurs. In the example of [22], one can change the Fano parameter by adjusting the phase of the leaked light. In addition, Fano resonances on plasmonic metamaterials [23], terahertz light-matter interaction [24] [25], and the study of polariton and exciton [26] are also fruitful applications for nanoscience, meta-material, polaritonics, and quantum computing. Also, the discussion of a quantum dot in an Aharonov-Bohm interferometer is another example that the Fano parameter $q$ should be defined as a complex number [27].

Furthermore, The Fano resonance in the context of thermodynamics was formulated as dissipative processes of a quantum system [28]. Regarding quantum chaos [29], there are emerging discussions on the Fano resonance. As one of the dissipative processes, the lifetime measurement of Fano resonance is started to discuss recently in the context of molecular science using the method of merged beam experiment [30]. In spectroscopy with ultrashort pulse laser such that attosecond laser [31] and XFEL [32], it is a recent known fact that the Fano parameter is affected by the pulse duration.

Similar considerations can often be made in various linear systems involving the superposition of continuum and a discrete states. Examples include control theory and signal processing. In those fields, a diagram called the Nyquist diagram (or Nyquist plot) is frequently used to confirm the stability of the system. This is plotted for each $\omega$ ($-\infty < \omega < \infty$) with the real part of the frequency response on the horizontal axis (In our discussion, it is the square of the amplitude of the mass 1) and the imaginary part of the frequency response on the vertical axis (Energy dissipation rate is expressed by it). There is a paper that plots the Nyquist diagram while changing the Fano parameter in the system where Fano resonance occurs [33].

## References

[1] Madden R P and Codling K 1965 *ApJ* **141** 364
[2] Fano U 1935 *Nuovo Cimento* **12** 154
[3] Fano U 1961 *Phys. Rev.* **124** 1866
[4] Fano U and Cooper J W 1968 *Rev. Mod. Phys.* **40** 441
[5] Fano U and Rau A R P 1986 *Atomic Collisions and Spectra* (Florida: Academic Press) p 230
[6] Kobayashi K, Aikawa H, Katsumoto S and Iye Y 2002 *Phys. Rev. Lett.* **88** 256806
[7] Riffe D M 2011 *Phys. Rev.* B **84** 064308
[8] Crawford F S Jr. 1968 *Waves* (*Berkeley Physics Course* Volume 3) (California: McGraw-Hill) p

116
[9] Joe Y S, Satanin A M and Kim C S 2006 *Phys. Scr.* **74** 259
[10] Satpathy S, Roy A and Mohapatra A 2012 *Eur. J. Phys.* **33** 863
[11] Lv B et al 2016 *Sci. Rep.* **6** 31884
[12] H Hanashiro et al 1979 *J. Phys. B: Atom. Mol. Phys.* **12** L775
[13] R J Slobodrian 1971 *Rep. Prog. Phys.* **34** 175
[14] Bransden B H and Joachain C J 2003 *Physics of Atoms and Molecules* (second ed.) (Essex: Pearson Education) p 593
[15] Huang L et al 2015 *AIP Adv.* **5** 017137
[16] Rau A R P 2004 *Phys. Scr.* **69** C10
[17] Boudouti E H El et al 2008 *J. Phys.: Condens. Matter* **20** 255212
[18] Hein S, Koch W and Nannen L 2010 *J. Fluid Mech.* **664** 238
[19] Amin M et al 2014 *USNC-URSI Radio Science Meeting (Joint with AP-S Symposium)* p 86
[20] Amoudache S et al 2016 *J. Appl. Phys.* **119** 114502
[21] Elayouch A et al 2016 *EPL* **116** 46004
[22] Yu Y et al 2014 *Appl. Phys. Lett.* **105** 061117
[23] Karmakar S et al 2020 *Opt. Lett.* **45** 3386
[24] Karmakar S et al 2019 *Phys. Status Solidi RRL* **13** 1900310
[25] Karmakar S et al 2020 *J. Phys. D: Appl. Phys.* **53** 415101
[26] Wang Y et al 2017 *Phys. Rev. Lett.* **118** 063602
[27] Kobayashi K et al 2002 *Phys. Rev. Lett.* **88** 256806
[28] Finkelstein-Shapiro D and Keller A 2018 *Phys. Rev. A* **97** 023411
[29] Augustovičová L D and Bohn J L 2018 *Phys. Rev. A* **98** 023419
[30] Blech A et al 2020 *Nat. Commun.* **11** 999
[31] Ott C et al 2013 *Science* **340** 716
[32] Aufleger L et al 2020 *J. Phys. B: At. Mol. Opt. Phys.* **53** 234002
[33] Kadri S, Fujiwara H and Sasaki K 2011 *Opt. Express* **19** 2317